\documentclass[useAMS,usenatbib]{mn2e}
\usepackage{natbib}
\usepackage{times}
\usepackage{rotate}
\usepackage{color}

\usepackage{graphicx}
\usepackage{amsmath,amssymb}

\newcommand{\ddd}{\displaystyle}
\newcommand{\Mpc}{\,\mathrm{Mpc}}

\newcommand{\BF}{\begin{figure}\begin{center}}
\newcommand{\EF}{\end{center}\end{figure}}
\newcommand{\BE}{\begin{equation}}
\newcommand{\EE}{\end{equation}}
\newcommand{\BEA}{\begin{eqnarray}}
\newcommand{\EEA}{\end{eqnarray}}

\newcommand{\simgt}{\lower.5ex\hbox{$\; \buildrel > \over \sim \;$}}
\newcommand{\simlt}{\lower.5ex\hbox{$\; \buildrel < \over \sim \;$}}

\begin{document}

\title{Constraints on small-scale cosmological fluctuations from
 SNe lensing dispersion}
\author[I. Ben-Dayan and R. Takahashi]
{Ido Ben-Dayan$^1$ and Ryuichi Takahashi$^2$
\\
$^{1}$Deutsches Elektronen-Synchrotron DESY, Theory Group, D-22603 Hamburg,
Germany \\
$^{2}$
Faculty of Science and Technology, Hirosaki University,
 3 Bunkyo-cho, Hirosaki, Aomori 036-8561, Japan
}

\date{\today, \, 
DESY 15-053
} 



\maketitle


\label{firstpage}

\begin{abstract}
We provide predictions on small-scale cosmological
 density power spectrum from supernova lensing dispersion.
Parameterizing the primordial power spectrum with running $\alpha$ and running
 of running $\beta$ of the spectral index, we exclude large positive 
 $\alpha$ and $\beta$ parameters which induce too large lensing dispersions
 over current observational upper bound.
We ran cosmological N-body simulations of collisionless dark matter particles
 to investigate non-linear evolution of the primordial power spectrum
 with positive running parameters.  
The initial small-scale enhancement of the power
 spectrum is largely erased when entering into the non-linear regime.
For example, even if the linear power spectrum at
 $k>10h {\rm Mpc}^{-1}$ is enhanced by $1-2$ orders of magnitude, 
 the enhancement much decreases to a factor of $2-3$ at late time
 ($z \leq 1.5$). 
Therefore, the lensing dispersion induced by the dark matter fluctuations
 weakly constrains the running parameters. 
When including baryon-cooling effects (which strongly enhance the 
 small-scale clustering), the constraint is comparable or tighter
 than the PLANCK constraint, depending on the UV cut-off.
Further investigations of the non-linear matter spectrum with
 baryonic processes is needed to reach a firm constraint.
\end{abstract}


\section{Introduction}
The plethora of cosmological observations has turned Cosmology into a
 quantitative science.
From combining several probes that observe the Universe at different epochs
 and have different systematics and statistics, emerged the
 `Concordance Model' of Cosmology, a six parameter model, most of them
 measured to the accuracy of a percent.   
 Among these probes, type Ia supernovae (SNe) are a powerful cosmological tool
 to directly
 measure the expansion history of the Universe
 \citep[see e.g. a recent review of][]{weinberg2013}.
The type Ia SNe are known as standard candles and one can measure the
 luminosity distances to them accurately.
Several SNe surveys have set strong constraints on cosmological models
 from the distance-redshift relation 
 \citep[e.g.][]{Riess1998,Perl1999,Riess2007,Sull2011,Camp2013}.
However the measured apparent magnitudes, have a residual scatter 
 arising from its intrinsic scatter and effects due to
 line-of-sight (LOS) structures.
The intrinsic scatter ($\sim 0.4$ mag) can be significantly reduced
 to $\sim 0.1$ mag by empirically calibrating the luminosity curves. 
The scatter due to photon deflection along the line of sight is composed of many different physical effects. The dominant effects are peculiar velocities at $z<0.1$ and gravitational lensing at $z\gtrsim0.3$. Within the realm of cosmological perturbation theory and the stochastic nature of the LOS structures, all effects are expressed as an integral over the power spectrum with appropriate kinematical factors \citep{BenDayan:2012wi}.
The lensing effects create residuals from the best-fitting curve
 in the magnitude-redshift relation. 
 The lensing magnifications of SNe can be extracted by correlating the
 residuals with the surface densities of nearby foreground galaxies.
 The lensing dispersion is roughly proportional to
 the SNe redshift $z$ as $\sim 0.06z$ mag \cite[e.g.][]{Holz2005,ido13}.
 
Detection of the lensing magnifications has been reported by
 several groups \citep{jonsson2006,jonsson2010,Kronborg2010,Karpenka2013,smith2014}.
J{\"o}nsson et al. (2010, hereafter J10) 
 used a sample of 175 SNe at $0.2 \lesssim z \lesssim 1$ from the SNLS 3yr data,
 and measured SNe lensing magnifications by correlating the SNe fluxes and
 its foreground nearby galaxies (modeled as singular isothermal spheres). 
Modelling the lensing dispersion as $\sigma_{\rm m}(z_{\rm s})=B z_{\rm s}$,
 they obtained a constraint
 $B \simeq 0.055^{+0.039}_{-0.041}$ mag ($1 \sigma$) and an upper bound
 $B \lesssim 0.095 (0.12)$ mag with $1(2)$-sigma confidence.  
More recently the Joint Lightcurve Analysis (JLA) \citep{Betoule:2014frx} 
 combined SNe from various observations to compile $740$ SNe 
for cosmological parameter inference, yielding $\Omega_{\rm m}=0.295\pm0.034$ very close to the PLANCK result \citep{Planck:2015xua}.
The JLA did the following errors analysis. First it considered the peculiar velocity dispersion, that is relevant only for very low redshift. In addition, it used J10 central value for the lensing dispersion. Last, it considered all additional possible sources of error and named it `coherent dispersion', $\sigma_{\rm coh}$. Adding all of these in quadrature produced the {\it total} dispersion, $\sigma_{\rm tot}(z\leq 1) \leq 0.12$ in magnitude. There is also a clear trend of the total dispersion decreasing with redshift, Fig. $7$ in JLA, suggesting that lensing might even be smaller than J10 central value. \footnote{ While the lensing dispersion grows with redshift, it is by no means the only contribution to the dispersion that can evolve with redshift. For example the SNe population can in principle evolve with redshift degrading the efficiency of SNe lensing as a probe. We thank Adam Riess for raising this issue.}  Throughout the paper we will take this number $\sigma_{\rm m}\leq 0.12$ as a conservative upper bound of the lensing dispersion, and $\sigma_{\rm m} \leq 0.095$ as a `plausible' upper bound, based on $1\sigma$ of J10. 

The photons' scatter {\it biases} the measurement. However, given a cosmological model, we can use gravitational lensing as a {\it probe} of Cosmology and Astrophysics.
There are several advantages of using SNe lensing as a cosmological probe. First, it has different systematics than other probes. Second, it is sensitive to different scales. For example, we shall see that lensing dispersion is mainly sensitive to $10^{-2}h {\rm Mpc}^{-1} \leq k \leq 10^3 h {\rm Mpc}^{-1}$. Third, even current data (or upper bound) is sufficient to constrain the cosmological parameters.
The gravitational lensing of type Ia SNe has been suggested as a probe of cosmological abundance of
 compact objects \citep[e.g.][]{Rauch1991,Holz1998,Holz2001,metcalf1999,
 mortsell2001,Minty:2001jg, Metcalf2007,Yoo2008} and for measuring the present amplitude of
 matter density fluctuations \citep[e.g.][]{Metcalf1999b,HF2000,Valageas2000,
 DV2006,Marra2013,QMA2014,Fedeli2014,IY2014}.
For instance,
 \cite{CQ2014} recently measured $\sigma_8$ (the present amplitude of density
 fluctuations at $8h^{-1} {\rm Mpc}$) using the non-Gaussianity of
 SNe magnitude distribution from $700$ SNe at $z<0.9$. 

Recently, \cite{Ido2014} used the lensing dispersion to constrain
 small-scale amplitude of primordial curvature perturbation.
They modelled the primordial power spectrum using running $\alpha\equiv dn_s/d \ln k$ and
 running of running $\beta\equiv d^2 n_s/d \ln k^2$ of the spectral index, and then discussed an 
 observational constraint on $\alpha$ and $\beta$ from the lensing dispersion.
 A subsequent work considered other parameterizations of the primordial power spectrum, \citep{Ido2014b}.
Current observations \citep{Planck:2015xua} measure $n_s(k_0) \simeq 0.96$, for the parameterization $P_{k}=A_s(k/k_0)^{n_s(k_0)-1}$, where $k_0$ is a suitable ``pivot scale''.
Focusing again on ``running", we consider: 
\begin{align}
n_s(k) = \ddd n_s(k_0)+\ddd\frac{\alpha(k_0)}{2} \ln\frac{k}{k_0}+\ddd\frac{\beta(k_0)}{6} \left( \ln \frac{k}{k_0} \right)^2 \,,
\end{align}
There is no compelling evidence for the more generalized form, but the best $\Delta \chi^2$, that takes into account the suppression of low multiples in the temperature anisotropies spectrum is with $\alpha=0, \beta=0.029$ \citep{Planck:2015xua}.\footnote{Non-trivial running is a \textit{prediction} if $\epsilon$ is non monotonic, see \cite{BenDayan:2009kv}, and a Higgs inflation version in \cite{Hamada:2014iga, Bezrukov:2014bra}.}	
 Given a single canonical scalar field $\phi$ and a potential $V(\phi)$, the above observables can be described by the slow-roll parameters, or the potential and its derivatives at the 'CMB point':
\BEA
n_s \hspace{-3mm} &=& \hspace{-3mm}  1+2\eta-6\epsilon\\
\alpha \hspace{-3mm} &=& \hspace{-3mm} 16\epsilon \eta-24 \epsilon^2-2\xi^2\\
\beta \hspace{-3mm} &=& \hspace{-3mm} -192 \epsilon^3+192 \epsilon^2\eta-32 \epsilon \eta^2-24\epsilon \xi^2+2\eta \xi^2+2\omega^3\\
\epsilon \hspace{-3mm} &=& \hspace{-3mm} \frac{1}{2}\left(\frac{V'}{V}\right)^2, \quad \eta=\frac{V''}{V}, \quad \xi^2=\frac{V'V'''}{V^2}, \quad \omega^3=\frac{V'^2V^{(4)}}{V^2} \nonumber
\EEA
where prime denotes derivative with respect to the inflaton. The best constraints on $\alpha,\beta$ with $k_0=0.05 {\rm Mpc^{-1}},\,n_s(k_0) \simeq 0.96$ are given by PLANCK \citep[]{planck2013xxii,Planck:2015xua}. 
 These analyses are only probing the range $H_0\leq k\lesssim 1 \Mpc^{-1}$. The lensing dispersion,
$\sigma_{\rm m}$ is sensitive to  $0.01 {\rm Mpc}^{-1} \lesssim k \lesssim 10^3 \Mpc^{-1}$, thus giving access to $2-3$ more decades of the spectrum. Hence, $\sigma_{\rm m}(z)$ is particularly sensitive to the quasi-linear and non-linear part of the spectrum. In terms of inflation, the direct measurement of $k\lesssim 1\Mpc^{-1}$ corresponds to about $8$ e-folds of inflation, leaving most of the power spectrum of $\sim 60$ e-folds out of reach. Hence, even after PLANCK there is still an enormous space of inflationary models allowed. It is therefore of crucial importance to infer as much of the spectrum as possible for a better inflationary model selection. The lensing dispersion constrains additional $4-7$ e-folds, yielding a total of $12-15$ e-folds.

The main limitation of the analysis in \citet{Ido2014} and \citet{Ido2014b}, was the knowledge of the non-linear power spectrum
$k\gtrsim 1 h {\rm Mpc^{-1}}$ at late times $z<2$, in the case of running $\alpha$ and running of running $\beta$.
In the absence of numerical simulations with proper initial conditions, enhancement of power on non-linear scales was parameterized by the enhancement that occurs in the Concordance Model (i.e. $\alpha=\beta=0$).
In this paper, we overcome the main limitation of previous works. 
We ran N-body simulations with non-zero $\alpha, \beta$ and derive a modified HaloFit formula for the non-linear power spectrum.
We calculate the lensing dispersion and compare it with
the observational upper bound $\sigma_{\rm tot}\leq 0.12$.
Then, large positive running parameters $\alpha$ and $\beta$ can be rejected.
We will also include the baryonic cooling effects, which strongly affect
 the small-scale clustering, in our investigation.

Before outlining the rest of the manuscript a valid question arises regarding the logic of considering just the lensing dispersion, rather than the full lensing probability distribution function (pdf). The calculated and experimentally quoted lensing dispersion is that of the distance modulus, $m\sim-2.5 \log_{10}F$. The flux $F$ is usually simplified to have a log-normal distribution \citep{Holz2005}. In that case, the distance modulus is a Gaussian random variable and all the information is contained in the dispersion, and one does not need to consider higher moments of the pdf. Consideration of higher moments is relevant for deviations from the log-normal distribution of the flux and is beyond he scope of this work. 

The paper is organized as follows.
In sec.2, we briefly describe the lensing dispersion formula.
In sec.3, we first present the primordial power spectrum parameterized with
 the running parameters ($\alpha$ and $\beta$), and then run the cosmological
 N-body simulations to follow its non-linear evolution accurately.
The fitting formula of the non-linear matter power spectra are presented in
 this section and in the Appendix.
Sec.4 presents our main results of the lensing dispersion and 
 observational constraints on the running parameters.
Here, we also include the baryon cooling effects.
 Conclusions and a discussion are in section 5.
Throughout this paper we adopt the concordance $\Lambda$CDM model,
 which is consistent with the PLANCK results \citep{planck2013xvi}.
The model is characterized by the matter density $\Omega_{\rm m} =0.3134$,
the baryon density $\Omega_{\rm b}=0.0486$, the cosmological constant density
$\Omega_\Lambda=0.6866$, the amplitude and the spectral index of
 primordial curvature power spectrum $A_{\rm s}=2.20 \times 10^{-9},\,n_{\rm s}=0.96$, and the Hubble expansion rate today
 $H_{0}=67.3~$km s$^{-1}$ Mpc$^{-1}$. We have verified that varying the other cosmological parameters does not change the conclusions beyond the quoted error bars  \citep[see][]{Ido2014}.

\section{Lensing dispersion}
When a distant SNe is gravitationally lensed by intervening
 structures, its flux is multiplied by a
 magnification factor $\mu$ and its apparent magnitude (or distance modulus)
 is changed by $\delta m = -2.5 \log_{10} \mu$.
 For a perturbed FLRW Universe, in the Poisson gauge, one starts with the line of sight (LOS) first order lensing contribution to the distance modulus,
\begin{align}
  \delta m (\eta_s^{(0)}) =\frac{5}{\ln 10} \int_{\eta_s^{(0)}}^{\eta_o}
  \frac{d\eta_1}{\Delta\eta}\frac{\eta_1 - \eta_s^{(0)}}{\eta_o - \eta_1}
  \Delta_2\Psi
 \,,
 \label{delmu}
\end{align}
where the gravitational potential $\Psi=\Psi(\eta_i,r_i,\tilde\theta^a)$ is evaluated along the past light-cone at $r_i=\eta_o-\eta_i$, $\eta_o$ is the observer conformal time, $\eta_s^{(0)}$ is the conformal time of the source with unperturbed geometry, $\Delta \eta(z)=\eta_o(z)-\eta_s^{(0)}(z)$ $=H_0^{-1} \int^z_0
 dy / \sqrt{\Omega_{\rm m}(1+y)^3+\Omega_{\Lambda}}$ and $\Delta_2$ is the 2D angular Laplacian, see \cite{BenDayan:2012wi,ido13} for technical terms and further explanations. Squaring (\ref{delmu}) and taking the ensemble average in Fourier space at a fixed observed redshift, gives the variance, $\sigma^2_{\rm m}(z)$.
The double LOS integral at $z\gtrsim0.3$, is dominated by the equal time part, and with $\mathrm{Si}(x \gg 1) \approx \pi/2$, gives:
\begin{align}
\label{sigmuLNL}
\sigma_{\rm m}^2 &\simeq \left(\frac{5}{ \ln 10}\right)^2
\frac{\pi}{\Delta\eta^2} \int_{\eta_s^{(0)}}^{\eta_o} d\eta_1 \int dk
 P_{\Psi}(k,\eta_1)k^2\\
 & ~~~~~~~~~~~~~~~~~~~~~~~~~~~~\times (\eta_1-\eta_s^{(0)})^2(\eta_o-\eta_1)^2\,,\nonumber
\end{align}
where $P_{\Psi}$ is the dimensionless power spectrum of
the \textit{gravitational potential}.
Hence, the lensing dispersion of supernovae is a direct measurement of the integrated late-time power spectrum. 
Let us stress that it is both UV and IR finite \citep{ido13}. 
In general, the $k^2$ enhancement makes $\sigma_{\rm m}$ a sensitive probe to the small scales of the power spectrum. This sensitivity is manifest if we switch to dimensionless variables, $\tilde \eta= H_0 \eta$ and $p=k/k_{\rm eq}$: 
 \begin{align}
\label{dimless}
\sigma_{\rm m}^2 \simeq \left(\frac{5}{ \ln 10}\right)^2 \frac{\pi}{\Delta \tilde \eta^2} \left(\frac{k_{\rm eq}}{H_0}\right)^3 \int d\tilde \eta_1 \int dp
P_{\Psi}(p,\tilde \eta_1)p^2\\
\times (\tilde \eta_1-\tilde \eta_s^{(0)})^2(\tilde \eta_o-\tilde \eta_1)^2\,.\nonumber
 \end{align}
It is easy to see that: $(a)$ The relevant physical scales are $H_0$ and $k_{\rm eq}$, which give an enhancement of $(k_{\rm eq}/H_0)^3$. $(b)$ The dispersion is really sensitive to scales smaller than the equality scale $p>1$. $(c)$ The nonlinear power spectrum (NLPS) has an additional redshift dependent physical scale which is the onset of nonlinearity $k_{\rm NL}$. For a given redshift, parameterizing the NLPS as $\sim C (k/k_{\rm NL})^{\nu},\, \nu<-3$, from some $k_{\rm NL}$, will have an additional parametric enhancement of $(k_{\rm NL}/k_{\rm eq})^3$ to the lensing dispersion.
Note that $\sigma_{\rm m}$ is in units of magnitude and the dispersion
 in equations (\ref{sigmuLNL}) and (\ref{dimless}) is \textit{exactly} the standard result in weak gravitational lensing \cite[e.g.][]{bs01}.
In principle, the lensing dispersion can probe the density fluctuations from a size of SNe at its maximum luminosity ($\approx 10^{15}$ cm)
 \citep[e.g.][]{metcalf1999,mortsell2001}
 to the present Horizon scale $H_0^{-1}$. From equation (\ref{dimless}), it is clear that large scales $k<k_{\rm eq}$ have a negligible contribution. At smaller scales, $k\sim1\, h {\rm Mpc}^{-1}$, perturbation theory breaks down. Additionally, we expect non-Gaussian effects to become relevant. Last, the Nyquist wave number of the simulations soon to be discussed is at most $k_{\rm Nyq}=64.3 h {\rm Mpc}^{-1}$. We therefore limit the integration at most to $=10^3 h {\rm Mpc}^{-1}$, while $k_{\rm min}=H_0$.

In the following sections, we will evaluate the non-linear matter power spectrum using cosmological N-body simulations.
Here, the (dimensional) matter power spectrum $P_{\rm m}$ and the (dimensionless) gravitational potential power spectrum $P_\Psi$
 are related via,
\BE
P_\Psi (k,z) = \frac{9}{8 \upi^2} H_0^4 \Omega_{\rm m}^2 \left( 1+z \right)^2 k^{-1} P_{\rm m} (k,z).
\label{potent}
\EE

\section{Non-linear matter power spectrum}

We run cosmological N-body simulations to investigate the non-linear matter
 power spectra of the various running spectral index models.
Our purpose is to obtain a fitting formula of the non-linear power
 spectra and calculate the lensing dispersion in equation (\ref{sigmuLNL}).

\subsection{Linear power spectra with running spectral indices}
\label{sec_linear_pk}

\begin{figure}
\begin{center}
\includegraphics[width=7.5cm]{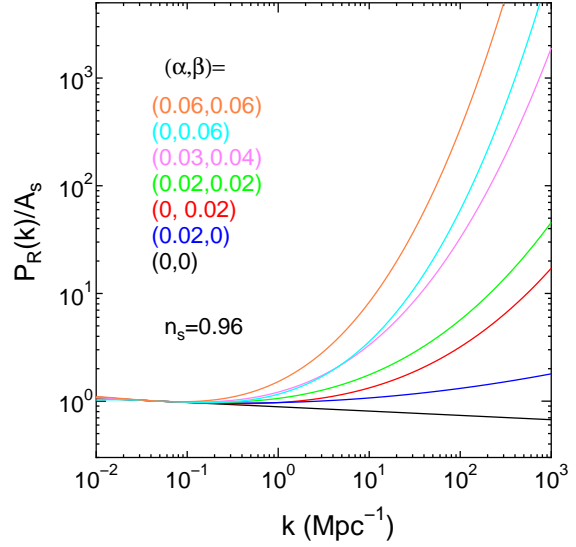}
\caption{Dimensionless curvature power spectrum $\mathcal{P}_\mathcal{R}(k)$
 in our running spectral index models:
 $(\alpha,\beta)=(0.06,0.06)$, $(0,0.06)$, $(0.03,0.04)$, $(0.02,0.02)$,
 $(0,0.02)$, $(0.02,0)$, $(0,0)$ from top to bottom.
Here, in vertical axis, the power spectrum is divided by its amplitude
 $A_{\rm s}$.
}
\label{fig_linear_pk}
\end{center}
\end{figure}

The dimensionless curvature power spectrum is given by
\BE
  \mathcal{P}_\mathcal{R}(k) = A_{\rm s} \left( \frac{k}{k_0}
 \right)^{n_{\rm s}-1+(\alpha/2) \ln (k/k_0) + (\beta/6) [\ln (k/k_0)]^2}, 
\label{eq_linear_pk}
\EE
where $k_0=0.05 {\rm Mpc}^{-1}$ is the PLANCK pivot scale, $A_{\rm s}=2.20
 \times 10^{-9}$ is the amplitude and $n_{\rm s}=0.96$ is the spectral index. 
For the running $\alpha$ and the running of running $\beta$ of the spectral
 index, we adopt seven models as
 $(\alpha,\beta)$ $=(0,0)$, $(0,0.02)$, $(0.02,0)$,
 $(0.02,0.02)$, $(0.03,0.04)$, $(0,0.06)$ and $(0.06,0.06)$.
Note that $A_{\rm s}$ is fixed for the running models ($\alpha \neq 0$,
 $\beta \neq 0$).
Then, the present mass variance $\sigma_8$ at $8h^{-1}$Mpc weakly
 increases as the running parameters ($\alpha$ and $\beta$) increase:
 $\sigma_8=0.829$ for $(\alpha,\beta)$ $=(0,0)$,
 $0.831$ for $(0,0.02)$, $0.833$ for $(0.02,0)$, $0.835$ for $(0.02,0.02)$,
 $0.839$ for (0.03,0.04),
 $0.834$ for $(0,0.06)$, $0.847$ for $(0.06,0.06)$, respectively.
Fig.\ref{fig_linear_pk} plots the curvature power spectrum in equation
 (\ref{eq_linear_pk}) in our running models.
The small-scale enhancement of $\mathcal{P}_\mathcal{R}(k)$ is
 clearly prominent.
Large $\alpha,\beta$ models strongly enhance $\mathcal{P}_\mathcal{R}(k)$
 by an order of magnitude for $k \gtrsim 10\,h {\rm Mpc}^{-1}$.
The PLANCK results already excluded large running models of
 $(\alpha,\beta)$ $=(0,0.06)$ and $(0.06,0.06)$ at $2 \sigma$
 level \citep{planck2013xxii}.
However, it is useful to study the general feature of the non-linear
 gravitational evolution with such strong small-scale enhancement. Furthermore, it provides an independent crosscheck, to the PLANCK analysis, and as we have mentioned, it probes scales beyond PLANCK's reach.

We prepare the linear matter power spectrum with the running parameters,
 $P_{\rm m,lin}(k;\alpha,\beta)$, as follows. 
We first evaluate the linear power spectrum without the running 
 $P_{\rm m,lin}(k;\alpha=0,\beta=0)$ using CAMB \citep{camb}. 
Then, simply multiplying a factor from equation (\ref{eq_linear_pk})
 to the output of CAMB as,
\BEA
 && \hspace{-0.5cm} P_{\rm m,lin}(k;\alpha,\beta) =
 P_{\rm m,lin}(k;\alpha=0,\beta=0)  \nonumber \\
 && \hspace{2.cm} \times \left( \frac{k}{k_0} \right)^{(\alpha/2) \ln (k/k_0)
 + (\beta/6) [\ln (k/k_0)]^2}.
\EEA

\subsection{Cosmological N-body simulations}

\begin{figure*}
\begin{center}
\includegraphics[width=17.cm]{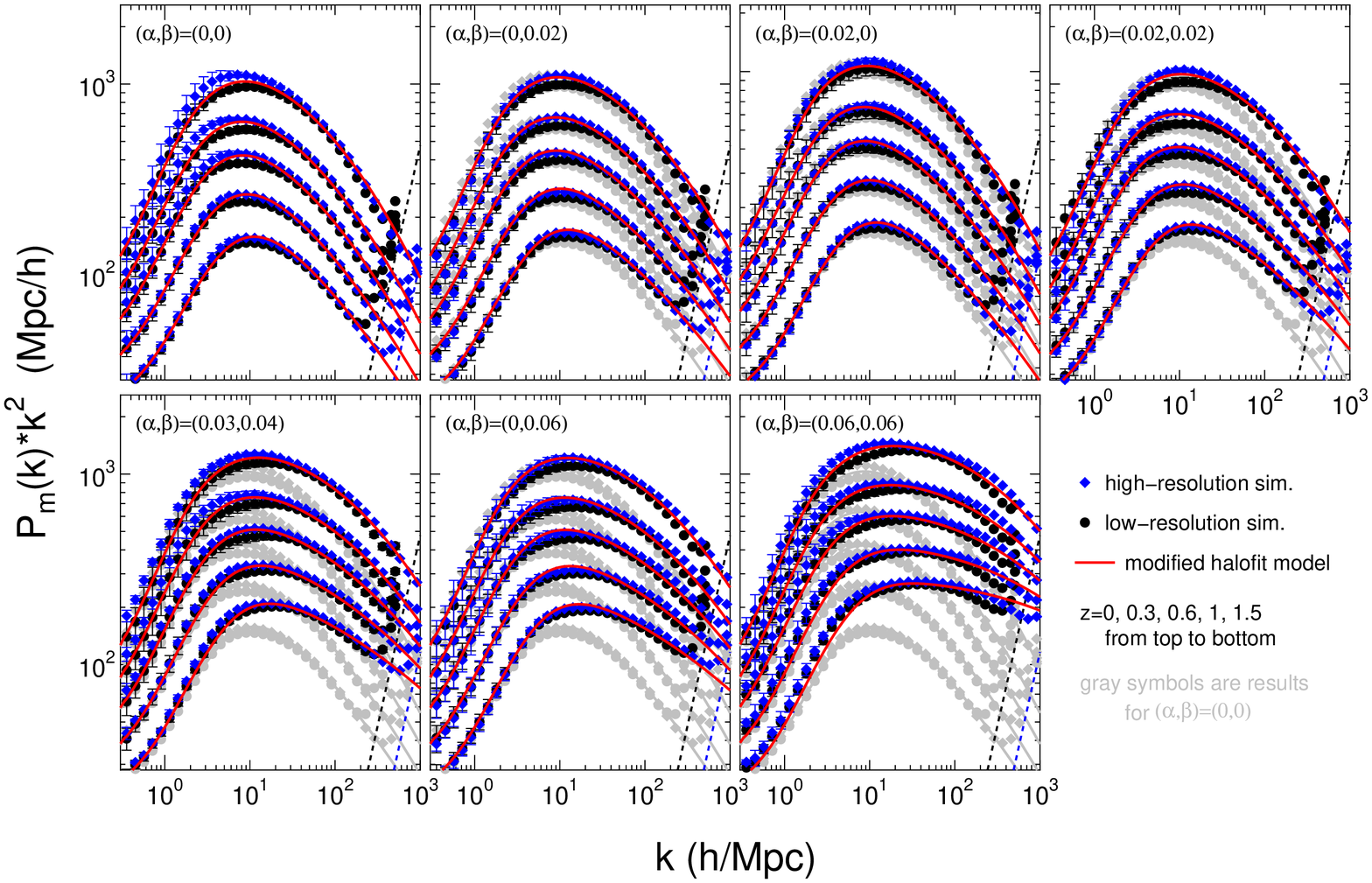}
\caption{Non-linear matter power spectra with running $\alpha$ and
 running of running $\beta$ of the spectral index.
The seven panels correspond to $(\alpha,\beta)$ $=(0,0)$, $(0,0.02)$,
 $(0.02,0)$, $(0.02,0.02)$, $(0.03,0.04)$, $(0,0.06)$ and $(0.06,0.06)$,
 respectively.
The redshifts are $z=0,0.3,0.6,1$ and $1.5$ from top to bottom in each panel.
In the vertical axis, the power spectrum is multiplied by a factor $k^2$ to
 show clearly our results.
The blue diamonds (black filled circles) are the high-(low-)resolution
 simulation results.
The dark gray symbols are the results for the no-running model
 $(\alpha,\beta)=(0,0)$.
The blue (black) dotted lines denote the shot noise in the
 high-(low-)resolution simulations.
The red curves are our modified halofit model (see main text and Appendix A
 for detailed discussion).
}
\label{fig_pk}
\end{center}
\end{figure*}

\begin{table*}
\hspace{-1.5cm}
\caption{Our Simulation Setting}
\setlength{\tabcolsep}{2pt}
\begin{tabular}{lcccccc}
\hline
\hline
   & $L(h^{-1} \rm{Mpc})$ & $N_{\rm p}^3$ &
 $k_{\rm Nyq} (h \rm{Mpc}^{-1})$ & $m_{\rm p} (h^{-1} M_\odot)$ &
 $z$ & $N_{\rm r}$  \\ 
\hline
 high-resolution & $100$ & $2048^3$ & $64.3$ & $1.0 \times 10^7$ & $0,0.3,0.6,1,1.5$ & $3$ \\
\hline
 low-resolution & $100$ & $1280^3$ & $40.2$ & $4.1 \times 10^7$ & $0,0.3,0.6,1,1.5$ & $6$ \\
\hline 
\label{table_sim}
\end{tabular}
\\
The different columns are the side length of simulation box $L$, the number of dark
 matter particles $N_{\rm p}^3$, the Nyquist wavenumber $k_{\rm Nyq}$,
 the particle mass $m_{\rm p}$, the redshifts of the simulation outputs $z$,
 and the number of realizations $N_{\rm r}$.
\end{table*}

We run cosmological numerical simulations of dark matter particles
 in cubic boxes with a side of $100h^{-1}$Mpc.
In order to check the numerical resolution, we run high- and low-resolution
 simulations in which the number of particles are $2048^3$ and $1280^3$.
The initial matter power spectrum is prepared following the procedure 
 in section \ref{sec_linear_pk}.
The initial positions and velocities of particles are given at redshift
 $z_{\rm init}=99$ based on second-order Lagrangian perturbation theory
 \citep{crocce2006,nishimichi2009}.

To follow the gravitational evolution of the dark matter particles,
 we employ a publicly available tree-PM code GreeM
 \citep{ishiyama2009,ishiyama2012}.
GreeM is tuned to accelerate the tree gravitational calculation, and it
 is fast especially in the strongly non-linear regime.
The PM meshes are set to be $1024^3 (640^3)$ and the particle Nyquist
 wavenumbers (which is determined by the mean particle separation)
 are $k_{\rm Nyq}=64.3 (40.2) h {\rm Mpc}^{-1}$ for the
 high-(low-)resolution simulations.
The small-scale enhancement seen in Fig.\ref{fig_linear_pk} is included
 up to the Nyquist wavenumber ($k<k_{\rm Nyq}$) in the initial conditions
 of our simulation.
In other words, the initial linear $P_{\rm lim}(k)$ is zero at $k>k_{\rm Nyq}$.  
The gravitational softening length is set to be $3\%$ of the mean particle
 separation.
The simulation snapshots are dumped at redshifts $z=0, 0.3, 0.6, 1$ and $1.5$.
We prepare $3 (6)$ independent realizations for the high-(low-)resolution
 simulations for each $(\alpha,\beta)$ model to reduce the sample variance.
Our simulation settings are summarized in Table \ref{table_sim}.

Fig.\ref{fig_pk} shows our simulation results of the matter power spectra
 with the various running parameters $(\alpha,\beta)$ at
 $z=0,0.3,0.6,1$ and $1.5$. 
The blue diamonds (black filled circles) with error bars are the mean power
 spectra with the standard deviations obtained from the realizations of
 high-(low-)resolution simulations.
The blue (black) dotted lines are the shot noise for the
 high-(low-)resolution simulations.
The high-resolution results are slightly higher (about $10\%$) than the
 low-resolution ones especially at low redshifts $z \leq 0.6$.
The initial $P_{\rm m,lin}(k)$ is exponentially enhanced for some large
 $\alpha$ and $\beta$, and hence the simulation results would strongly
 depend on the resolution especially at wavenumbers comparable to or
 larger than the Nyquist wavenumber. 
However the differences between the high- and low-resolution results
 are smaller than $20 \%$ for $k=3-300h {\rm Mpc}^{-1}$ at $z=0-1.5$.
The red solid curves are our fitting formula based on the
 Halofit model for the $\Lambda$CDM model \citep{smith2003,takahashi2012},
 but slightly modified for the running spectral index models.
The Halofit agrees with the simulation results within $18.3 (28.3) \%$ relative
 error for $k<300 (600) h {\rm Mpc}^{-1}$ at $z=0-1.5$.
Details of the fitting parameters are given in Appendix A.

As seen in the figure, the matter power spectrum $P_{\rm m}(k)$ with 
 larger running parameters is further enhanced at small scales.
For example, $P_{\rm m}(k)$ with $(\alpha,\beta)=(0.06,0.06)$ and
 $(0.03,0.04)$ are about $3$ and $2$ times larger than that without the
 running at $k=50h{\rm Mpc}^{-1}$. 
However, the enhancement is much less prominent than those seen in the
 linear power spectrum (see Fig.\ref{fig_linear_pk}) which shows $1-2$
 orders of magnitude enhancement at $k>10h{\rm Mpc}^{-1}$.
Our numerical results show that the small-scale enhancement in the initial
 $P_{\rm m}(k)$ are largely smeared out at late times. 
This result exhibits power transfers from large to small scales,
via mode coupling, and thus suggests that the initial large-scale power mainly determines
the non-linear small-scale power at late times.
These power transfers in the non-linear gravitational evolution has been
 well studied for the models in which an initial $P_{\rm m}(k)$ has a peak
 (like a Gaussian) at a specific scale
 \citep[e.g.,][]{bp1997,pr2006,bp2009,ny2013,nishimichi2014} and in models where the initial
 $P_{\rm m}(k)$ is damped at small scale such as warm
 dark matter models \citep[e.g.,][]{lwp1991,wc2000,
 sm2011,viel2012,itti2014}.
Such a smearing of small-scale enhancement makes it
 difficult to discriminate the running parameters
 $(\alpha,\beta)$ from the nonlinear $P_{\rm m}(k)$.

Before closing this section, let us comment on effects of the primordial
 small-scale enhancement on a halo mass function.
We compute the mass function by identifying haloes in the low-resolution
 simulations using standard friends-of-friends algorithm with linking length
 of $0.17$ times the mean particle separation.
Then, as expected, the strong running models predict more numerous
 small halos, but the differences of the mass functions among the seven
 models are only less than a factor $2$ at $z=0-1.5$.
For example, even for the smallest halo of $4 \times 10^9 h^{-1} M_{\odot}$
 (which contains $100$ simulation particles), the largest running model
 ($\alpha=\beta=0.06$) has about $2 (1.4)$ times more haloes comparing to
 the no running model ($\alpha=\beta=0$) at $z=1.5 (0)$.
This result suggests that the primordial small-scale enhancement does not
 significantly affect the halo mass function at late time ($z \leq 1.5$).

\section{Results}

This section presents our main results of an observational upper bound on
 the running parameters $(\alpha,\beta)$ from the lensing dispersion.
First, subsection \ref{dm_only} gives the results based on
 the non-linear matter power spectra of dark matter fluctuations obtained
 in previous section.
Then, subsection \ref{baryons} includes baryonic
 effects on the matter power spectra and the lensing dispersion.
We will show that the baryonic effects strongly enhance the lensing
 dispersion and tighten the upper bound on $\alpha$ and $\beta$.

\subsection{Dark Matter Only}
\label{dm_only}

\begin{figure}
\begin{center}
\includegraphics[width=8.cm]{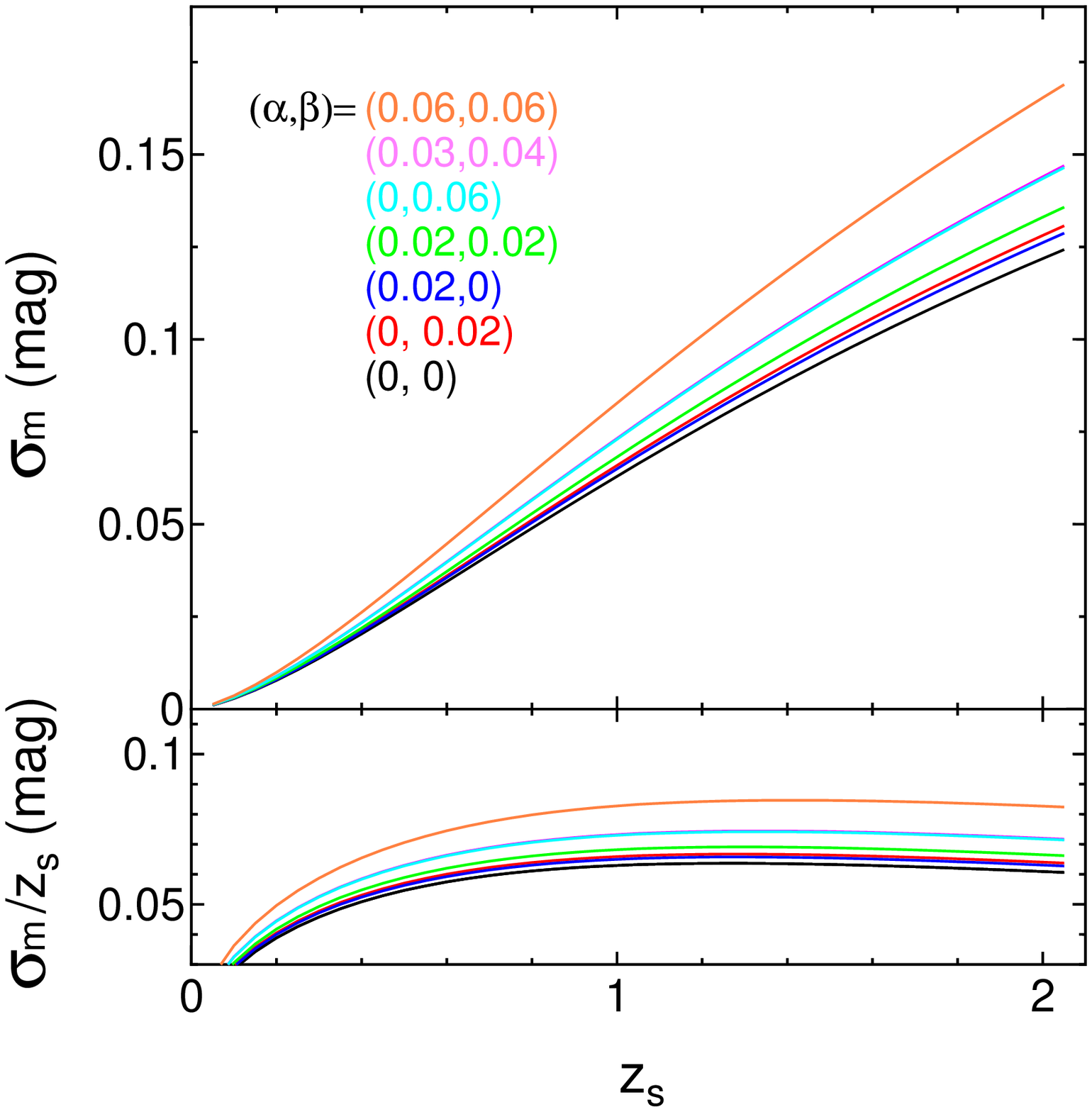}
\caption{Lensing dispersion as a function of source redshift $z_{\rm s}$ for
 the various running $\alpha$ and running of running $\beta$
 of the spectral index.
Top panel is the lensing dispersion in units of magnitude, and bottom one is
 the lensing dispersion divided by the source redshift.}
\label{fig_lens_disp}
\end{center}
\end{figure}

\begin{figure}
\begin{center}
\includegraphics[width=11.cm]{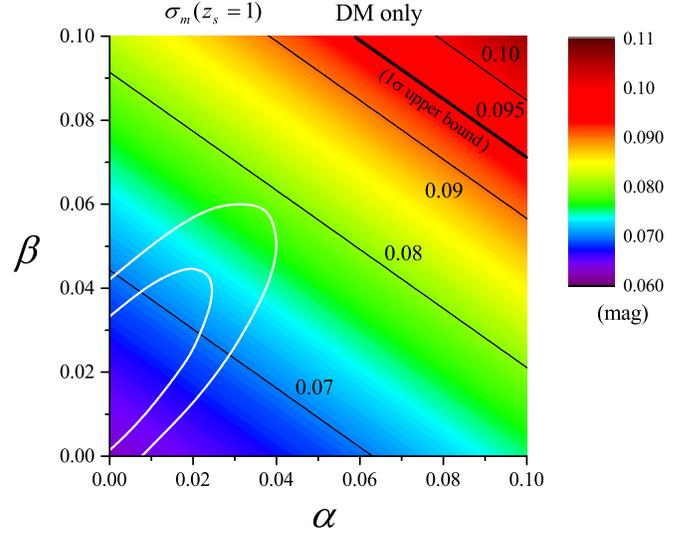}
\caption{Contour plot of the lensing dispersion at $z_{\rm s}=1$ in
 $\alpha \beta$ plane.
The black solid lines correspond to $\sigma_{\rm m}=0.07,0.08,0.09, \cdots$
 in steps of $0.01$ (mag).
 The thick line ($\sigma_{\rm m}=0.095$) denotes the observational upper bound
 (1 sigma) from the Supernova Legacy Survey (J10).
The white contours are the PLANCK \citep{planck2013xxii} constraints of 1 and 2 sigma regions.
}
\label{fig_contour}
\end{center}
\end{figure}

Fig.\ref{fig_lens_disp} shows our results of the lensing dispersion for 
 the various running parameters $(\alpha,\beta)$.
Here, we numerically calculate the lensing dispersion in equation
 (\ref{sigmuLNL}) using our fitting formula of $P_{\rm m}(k)$.
The figure clearly shows the lensing dispersion strongly depends on the
 running parameters.
For instance, the lensing dispersion for the strongest running model
 ($\alpha=\beta=0.06$) is about $40\%$ higher than the
 Concordance Model $(\alpha=\beta=0)$ at $0 < z_{\rm s} \leq 2$.
The bottom panel is the lensing dispersion divided by the source redshift,
 which shows $\sigma_{\rm m}(z_{\rm s})$ is roughly proportional to
 $z_{\rm s}$ at $z_{\rm s} \gtrsim 0.8$ \cite[e.g.][]{Holz2005,ido13}.

We comment on the interval of the integration of the lensing dispersion.
The integrand is proportional to
 $P_{\rm m}(k) k^2$ per $d \ln k$ from equations (\ref{sigmuLNL}) and (\ref{potent}),
 and thus the contribution to the integral comes from the peak of
 $P_{\rm m}(k) k^2$ which corresponds to
 $k \approx 10 h {\rm Mpc}^{-1}$ from Fig.\ref{fig_pk}.
Thus, the lensing dispersion does not strongly depend on the interval of
 the integral, if $k_{\rm min} (k_{\rm max})$ is much smaller (larger) than
 $\approx 10h{\rm Mpc}^{-1}$.
For example, if we set the maximum wavenumber $k_{\rm max}=10^{10} h
 {\rm Mpc}^{-1}$, instead of $10^{3} h {\rm Mpc}^{-1}$,
 the lensing dispersion becomes $6\%$ larger for
 $\alpha=\beta=0.06$ and less than $3\%$ larger for the other models
 at $z_{\rm s}=1$.

Fig.\ref{fig_contour} is a contour plot of the lensing dispersion at
 $z_{\rm s}=1$ in $\alpha \beta$ plane.
Here we apply our fitting formula of $P_{\rm m}(k)$ in a range of
 $0<\alpha,\beta<0.1$.
The thin black lines correspond to $\sigma_{\rm m}=0.07,0.08,0.09, \cdots$
 (mag).
The thick line ($\sigma_{\rm m}=0.095$) corresponds to the 
 observational upper bound ($68\%$ C.L.) from the Supernova Legacy
 Survey (SNLS, J10).
White contours are PLANCK constraints of 1 and 2 sigma
 regions \citep[Planck+WP+BAO in][]{planck2013xxii}.
The dependence of $\alpha,\beta$ in $\sigma_{\rm m}$ (seen in the solid
 lines) comes from the fitting formula (\ref{eq_f_p}),
 $3.32 \alpha + 4.72 \beta =$ const. 
Then, the SNLS constraint can be rewritten as $\alpha+1.42 \beta<0.227$
 ($95\%$ CL), and  $\alpha+1.42 \beta<0.167$ for the 'plausible upper bound'  ($68\%$ CL).
Our constraint may seem weaker than the PLANCK results, but it is important to note,
that unlike CMB results, ours extend
all the way to $k_{\rm max}=10^3 h {\rm Mpc}^{-1}$. So while the result is not compelling as one may
have hoped for, it still contains new information about the small scales of the primordial power
spectrum in that it extends PLANCK constraints to the $k_{\rm max}$ scale, if lensing is actually observed.

Our numerical results of the lensing dispersion at
 $0.6 \leq z_{\rm s} \leq 2$ and $0 \leq \alpha,\beta \leq 0.2$ can be
 fitted as,
\BE
 \frac{\sigma_{\rm m}(z_{\rm s};\alpha,\beta)}{z_{\rm s}} = f_0(z_{\rm s})
 + f_1(z_{\rm s}) \left\{ x + f_2(z_{\rm s}) x^{3.4} \right\}^{0.98}
 ~~{\rm mag}
\EE
where $x=3.32 \alpha+4.72 \beta$ and
\BEA
 && f_0(z_{\rm s})=0.061 + 0.010 (z_{\rm s}-1) -0.012 (z_{\rm s}-1)^2,
 \nonumber \\
 && f_1(z_{\rm s})=0.038 + 0.003 (z_{\rm s}-1),
 \nonumber \\
 && f_2(z_{\rm s})=0.75 + 0.37 (z_{\rm s}-1) -0.20 (z_{\rm s}-1)^2.
\EEA
The fitting function can reproduce the our
 numerical results within a relative error of $4.3 \%$ for
 $\sigma_{\rm m}/z_{\rm s}<0.3$mag.
Since $\sigma_{\rm m}$ is roughly proportional to $z_{\rm s}$, 
 $z_{\rm s}$-dependencies of the functions $f_{0,1,2}(z_{\rm s})$ are
 relatively weak.

\subsection{Including Baryonic Effects}
\label{baryons}

\begin{figure}
\begin{center}
\includegraphics[width=7.cm]{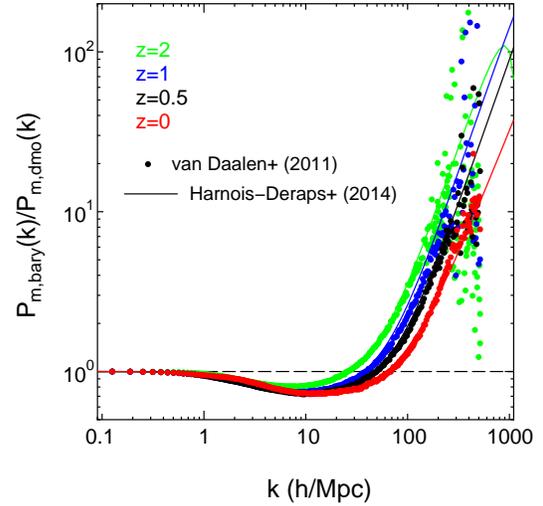}
\caption{The baryon feedback bias in equation (\ref{baryon_feedback_bias})
 at $z=0,0.5,1$ and $2$.
The vertical axis is the power spectra with baryons divided by power spectra with
 dark matter only.
The dots are simulation results of \citet{vanDaalen2011} and the solid
 curves are the fitting formula of \citet{HD2014}.}
\label{fig_baryon_bias}
\end{center}
\end{figure}

\begin{figure}
\begin{center}
\includegraphics[width=7.cm]{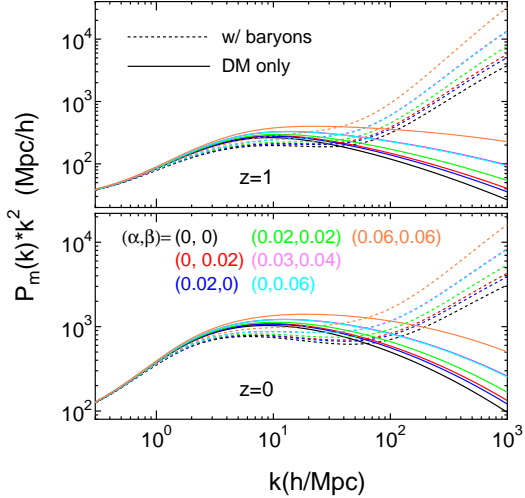}
\caption{Matter power spectra with baryons (dotted curves) and without it
 (solid curves) at $z=0$ and $1$.
The solid curves are our Halofit results, and the dotted curves are
 products of the baryon bias factor (\ref{baryon_feedback_bias})
 times our Halofit results.}
\label{fig_pk_baryon}
\end{center}
\end{figure}

In this subsection, we discuss the baryon-cooling effects on the matter
 power spectrum and the lensing dispersion.
In our analysis, we adopt previous results of high-resolution hydrodynamic simulations
 to incorporate the baryonic effects into the lensing dispersion calculation.
The baryon cooling naively enhances the matter power spectrum by some tens of
 percent at $k = 10h {\rm Mpc}^{-1}$ and
 the enhancement is more significant for smaller scales
 \cite[e.g.][]{jing2006,rudd2008,casarini2011}.
Recently, two groups of the OverWhelmingly Large
 Simulations (OWLS) project \citep{schaye2010,vanDaalen2011} and the
 Illustris project \citep{vogels2014n,vogels2014} suggested that
 the AGN feedback suppressed $P_{\rm m}(k)$ at intermediate scale
 ($k=1-10h {\rm Mpc}^{-1}$) but the baryon cooling strongly enhanced
 $P_{\rm m}(k)$ by orders of magnitude at small scales
 ($k \gtrsim 100h {\rm Mpc}^{-1}$).
These previous numerical results are roughly consistent but slightly
 different from one another, since it strongly depends
 on modelling of complicated astrophysical process such as
 star formation, radiative cooling, supernova and AGN feedback.
In this paper, we use the results of the most realistic model
 (``AGN$\_$WMAP7'') in the OWLS project.
They used extended version of Gadget3 \citep{springel2005} which used
 the treePM algorithm to calculate the gravitational evolution and
 smoothed particle hydrodynamics to follow the evolution of gas particles.
They employed $512^3$ CDM particles and equal number of baryonic particles
 in a cubic box of $100h {\rm Mpc}^{-1}$ on a side. 
Their Nyquist wavenumber is $16h {\rm Mpc}^{-1}$ and their softening
 length ($r_{\rm soft} = 7.8 h^{-1}$kpc) corresponds to a wavenumber
 $2 \pi / r_{\rm soft} \simeq 800h {\rm Mpc}^{-1}$.
They provided tabulated data of the matter power spectra up to
 $k \simeq 500h {\rm Mpc}^{-1}$
 on their website \footnote{http://vd11.strw.leidenuniv.nl/}.
Very recently, \cite{HD2014} made a useful fitting formula of the matter
 power spectra with baryons obtained by the OWLS project.
They fitted the ratio of the power spectra with baryons $P_{\rm m,bary}(k,z)$
 to that with dark matter only $P_{\rm m,dmo}(k,z)$, called
 ``the baryon feedback bias'', which is defined as,
\BE
 b^2_{\rm bary}(k,z) = \frac{P_{\rm m, bary}(k,z)}{P_{\rm m, dmo}(k,z)}.
\label{baryon_feedback_bias}
\EE
Fig.\ref{fig_baryon_bias} plots the baryon feedback bias in
 Eq.(\ref{baryon_feedback_bias}) as a function of $k$ at $z=0,0.5,1$ and $2$.
As clearly seen in the figure, the AGN feedback suppresses the matter
 power spectrum at $k \approx 10h {\rm Mpc}^{-1}$ at some tens of percent,
 but the baryon cooling strongly enhances it at
 $k \gtrsim 100h {\rm Mpc}^{-1}$ by an order of magnitude.
The fitting formula can reproduce the numerical results within $5\%$ relative
 error for $k<100h {\rm Mpc}^{-1}$ at $z<1.5$ \citep{HD2014}.

\begin{figure*}
\begin{center}
\includegraphics[width=16.cm]{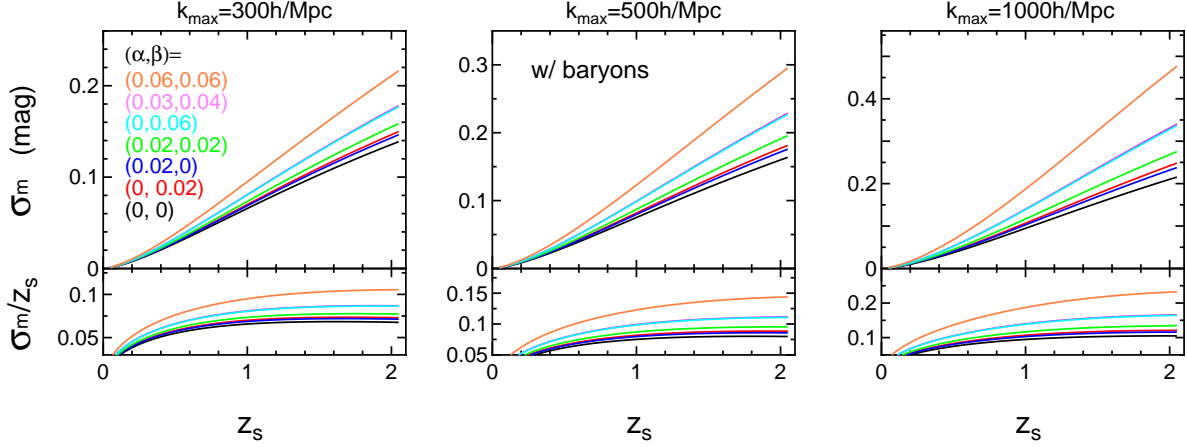}
\caption{Same as Fig.\ref{fig_lens_disp}, but including baryonic effects on
 the matter power spectrum. 
Each panel corresponds to various maximum wave numbers in the integral of
 the lensing dispersion (\ref{sigmuLNL}), $k_{\rm max}=300, 500$ and
 $1000\,h {\rm Mpc}^{-1}$ from left to right.
The baryons clearly enhance the lensing dispersion compared to
 the previous results in Fig.\ref{fig_lens_disp}.
}
\label{fig_lens_disp_baryon}
\end{center}
\end{figure*}

\begin{figure*}
\begin{center}
\includegraphics[width=18.cm]{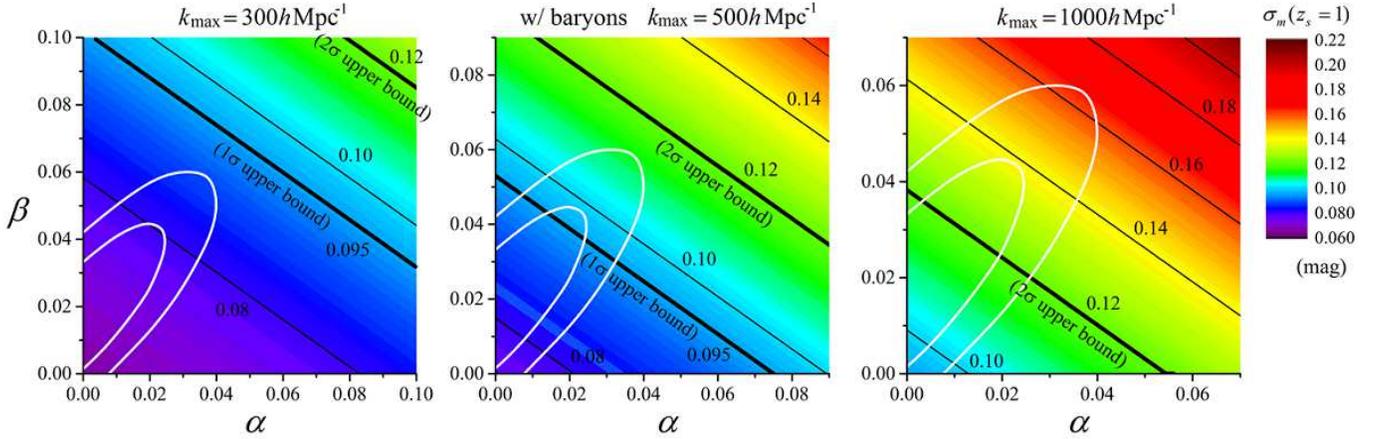}
\caption{Same as Fig.\ref{fig_contour}, but including baryonic effects on
 the matter power spectra. Each panel is for $k_{\rm max}=300, 500$ and
 $1000\,h {\rm Mpc}^{-1}$ from left to right.
 Note again that the thick solid lines correspond to 1- and 2-sigma observational upper bounds from J10 and JLA, while the white contours are 
the PLANCK \citep{planck2013xxii} 1 and 2 sigma likelihood contours.
}
\label{fig_contours_baryon}
\end{center}
\end{figure*}

In this paper, we simply multiply the bias factor (\ref{baryon_feedback_bias})
 to the results $P_{\rm m}(k)$ of our dark matter simulations to include
 the baryonic effects.
The bias factor was obtained for WMAP 7yr
 cosmological model without the running spectral index \citep{vanDaalen2011}.
However, we assume that the baryon feedback bias does not
 depend on the running parameters $\alpha,\beta$.
Positive running $\alpha,\beta$ model probably enhances formations of
 small-scale clumps (such as dwarf galaxies and star clusters) at high
 redshift, which would affect the non-linear power spectrum.
However, the dependence of the bias on $\alpha,\beta$ is unknown at present
 and further investigations (hydrodynamical simulations with running spectral
 index) are necessary.
Fig.\ref{fig_pk_baryon} shows the power spectra with baryons (dotted curves) 
 and without it (solid curves) at $z=0$ (lower panel) and $z=1$ (upper one).
In this figure, $P_{\rm m,bary}(k) k^2$ is a monotonically increasing
 function for large $k (\gtrsim 100h {\rm Mpc}^{-1})$ \footnote{The original result of $P_{\rm m,bary}(k,z) k^2$ in the OWLS project 
 is not an increasing function of $k$, rather roughly constant at
 $k \gtrsim 30 h {\rm Mpc}^{-1}$.
Their dark-matter simulation result $P_{\rm m,dmo}(k,z)$ is slightly smaller
 than ours at small scales ($k \gtrsim 100h {\rm Mpc}^{-1}$),
 which results in the difference of $P_{\rm m,bary}(k,z)$.}.
Thus, the integration in the lensing dispersion (\ref{sigmuLNL})
 strongly depends on $k_{\rm max}$ (it does not formally converge
 at $k_{\rm max} \rightarrow \infty$).
In what follows, we will present the lensing dispersion results for several
 $k_{\rm max}$.

Fig.\ref{fig_lens_disp_baryon} shows the lensing dispersions with the baryons.
Three panels are for various maximum wavenumbers in the integration
 (\ref{sigmuLNL}), $k_{\rm max}=300, 500$ and $1000h {\rm Mpc}^{-1}$
 (where the maximum $k_{\rm max}$ is set to be $1000h {\rm Mpc}^{-1}$, because
 the fitting formula of the bias factor $b^2_{\rm bary}$ becomes negative for
 $k>1000h{\rm Mpc}^{-1}$ at high redshift $z > 2$ \citep{HD2014}).
As expected, the results strongly depend on $k_{\rm max}$.
Depending on the chosen $k_{max}$ various values of $\alpha,\beta$ are excluded, by the conservative (plausible) upper bound $\sigma_{m}(z\leq 1) \leq 0.12 \, (0.095)$. 

Fig.\ref{fig_contours_baryon} is the contour plot of the lensing dispersion
 with the baryons at $z_{\rm s}=1$.
The result $\sigma_{\rm m}$ for $k_{\rm max}=300h {\rm Mpc}^{-1}$ is
 similar to the previous result in Fig.\ref{fig_contour}, but $\sigma_{\rm m}$
 is much larger than the previous results for
 $k_{\rm max}=500$ and $1000\,h{\rm Mpc^{-1}}$.
The SNLS 1-(2-)sigma constraints can be rewritten as
$\alpha+ 1.42\beta < 0.145 (0.222)$, $0.075 (0.139)$ and $0.0006 (0.054)$ 
 for $k_{\rm max}=300,500$ and $1000h{\rm Mpc^{-1}}$, respectively. 
 Our numerical results of $\sigma_{\rm m}(z_{\rm s};\alpha,\beta,k_{\rm max})$
 at $0.6 \leq z_{\rm s} \leq 2$ and $100 h {\rm Mpc}^{-1} \leq k_{\rm max}
 \leq 1000 h {\rm Mpc}^{-1}$ can be fitted as,
\BE
 \frac{\sigma_{\rm m}(z_{\rm s};\alpha,\beta,k_{\rm max})}{z_{\rm s}} =
 f_0(z_{\rm s},k_{\rm max}) \, {\rm e}^{f_1(z_{\rm s},k_{\rm max}) x
 + 1.8 x^2} ~~{\rm mag}
\EE
where $x=3.32 \alpha+4.72 \beta$ and
\BEA
 && f_0(z_{\rm s},k_{\rm max}) = \left[ 0.051 + 0.009 (z_{\rm s}-1) -
 0.009 (z_{\rm s}-1)^2 \right] \nonumber \\
 && \hspace{2.4cm} \times \left( 1+0.09 k_{{\rm max},100} \right),
 \nonumber \\
 && f_1(z_{\rm s},k_{\rm max})= \left[ 0.046 + 0.017 (z_{\rm s}-1) \right]
 \nonumber \\
 && \hspace{2.4cm} \times \left( 1+5.38 k_{{\rm max},100}
 - 0.27 k_{{\rm max},100}^2 \right), \nonumber \\
\EEA
where $k_{{\rm max},100}=(k_{\rm max}/100 h {\rm Mpc}^{-1})$. 
The fitting function agrees with our numerical results within a relative
 error of $8.7 \%$ for $0 \leq \alpha,\beta \leq 0.1$.
A byproduct of this analysis, is an easy derivation of the validity of
the baryon feedback. Taking $\alpha=\beta=0$, we hit the dispersion upper
bound of $\sigma(z\leq 1)\leq 0.12$ if we integrate up to
$k_{\rm max}=1500 h {\rm Mpc^{-1}}$, while the `plausible' $\sigma<0.095$ is
reached for $k_{\rm max}=950 h {\rm Mpc^{-1}}$. Hence the baryon bias has to be significantly modified and/or non-Gaussian effects have to be accounted for to explain the absence of detection of SNe lensing.

\section{Conclusions and Discussion}

We discussed the observational constraints on the small-scale power 
 spectrum of the primordial curvature perturbation from SNe lensing dispersion.
We modelled the common initial power spectrum with running parameters
 $\alpha$ and $\beta$ of the spectral index, and
 ran the cosmological high-resolution N-body simulations to follow its
 non-linear evolution.
We found that the small-scale enhancement in the primordial power spectrum
 is largely erased at late times during non-linear evolution, which suggests that
 it would be difficult to probe the small-scale primordial fluctuations
 using late time measurements. 
For the dark matter models, the observational constraints on $\alpha$ and
 $\beta$ from the current observational data of the SNLS of JLA are
 weaker than the PLANCK results.
However, if we include the baryonic effects, which strongly enhance the
 non-linear matter power spectrum at small-scales
 ($k \gtrsim 100h {\rm Mpc}$), the constraints are comparable to or tighter
 than PLANCK.

Let us discuss the accuracy of our Halofit formula in the high-$k$ limit. 
The fitting formula was made to fit the simulation results
in a finite range of $2h {\rm Mpc}^{-1}$ $< k <$ $600 h {\rm Mpc}^{-1}$
(see Appendix A).
In the high-$k$ limit, the simulation resolution limits the accuracy.
In the initial particle distribution, the linear $P_{\rm m}(k)$ is included up to the
Nyquist wavenumber $k_{\rm Nyq}$ and is exactly zero for $k>k_{\rm Nyq}$, and thus our
simulation results may underestimate the non-linear
 $P_{\rm m}(k)$ for $k>k_{\rm Nyq}$ at late times. Evidence for such an underestimation is the enhanced power spectrum in the high resolution simulations compared to the low resolution ones ($k_{\rm Nyq}=64$ and $40\, h {\rm Mpc}^{-1}$, respectively). Additional hint for such an underestimation is that at some point the nonlinear power spectrum becomes smaller than the linear one for very large wavenumbers. For example at $k\simeq 1000 \,h{\rm Mpc}^{-1}$ for $z=0,\,\alpha=\beta=0.03$ and at $k\simeq 120 \,h{\rm Mpc}^{-1}$ for $z=0,\,\alpha=\beta=0.06$.  
This lack of resolution would degrade the accuracy at small-scales especially
 for large $\alpha, \beta$ models.
However, as seen in Fig.\ref{fig_pk}, the high- and low-resolution simulations with aforementioned 
Nyquist wave numbers, 
give similar results with less than $20 \%$ difference for $k<300\,h{\rm Mpc}^{-1}$,
suggesting the safe convergence of our simulations. 
Further studies investigating the non-linear $P_{\rm m}(k)$ with positive $\alpha,\beta$
at such smaller-scales are necessary.

As we have seen, baryonic effects strongly influence our conclusions.
Baryonic simulation results strongly depend on assumptions about various
astrophysical processes (star formation and AGN feedback, stellar mass function and so on).
Our results are based on two assumptions:
i) We have used the \cite{vanDaalen2011} results.
  (In other words, we assume that the simulation of \cite{vanDaalen2011} is correct.)
ii) The baryon feedback bias in \eqref{baryon_feedback_bias}
 can be used even with large $\alpha$ and $\beta$.
Then, under these assumptions, we can safely set the strong constraints.
It may very well be that assumption ii) is incorrect, and simulations of baryons with initial $\alpha$ and $\beta$ will produce a better behaved spectrum, similar to our dark matter simulation results.
If our assumptions will be verified, it is clear that baryons have an overwhelming effect, as they induce $\mathcal{O} (1)$ changes of the power spectrum at $k\gtrsim 5\,h {\rm Mpc}^{-1}$. Furthermore, the changes in the power spectrum are non-monotonic in $k$. Last, with current simulation data, baryon cooling destroys the UV convergence, $P_{\rm m}(k) k^2$ diverges at
 $k \rightarrow \infty$, making the lensing dispersion a UV sensitive quantity. It is therefore senseless to derive conclusions on the power spectrum without the inclusion of baryons. Moreover, any (late-time) astrophysical probe that is sensitive to such scales, requires an accurate baryon modelling before deriving any conclusions about cosmology. 
As discussed in sec.\ref{baryons}, we need high-resolution hydrodynamic
 simulations with various initial conditions, that can probe very small-scales to reach a firm conclusion.
Given the huge effect of baryons on lensing dispersion, SNe lensing might be a useful calibrator for 
 baryonic processes. Especially since inclusion of baryonic effects in the Concordance Model ($\alpha=\beta=0$), implies a diverging power spectrum, that is unphysical, and gives a lensing dispersion very close to the observational bound for $k_{max}\sim 1000 ,h {\rm Mpc}^{-1}$.

Another natural development is considering higher moments of the lensing pdf.
Including small-scale fluctuations, the probability distribution of the
 magnification would be strongly non-Gaussian (highly skewed).
In fact, according to previous ray-tracing simulations in standard
 $\Lambda$CDM universe \citep[e.g.][]{hilb07,taka11b},
 the magnification PDF is rather a (modified) lognormal distribution with
 power-law tail at high $\mu$. 
Hence, the simple dispersion would not be sufficient to
 describe the lensing effects, and then the higher-order statistics such
 as skewness and kurtosis provide independent useful information. 

Finally, let us comment on future constraints on the running parameters
 $\alpha$ and $\beta$, and on the power spectrum at small scales.
Current tightest constraint on $\alpha$ and
 $\beta$ comes from PLANCK, which constrains them within $O(0.01)$ accuracy.
Adding to the CMB observations, future radio telescopes measuring 21cm
 fluctuations such as the Square Kilometer
 Array\footnote{see https://www.skatelescope.org/} and the
 Omniscope\footnote{http://space.mit.edu/instrumentation/omniscope}
 will constrain them within $O(0.001)$ accuracy
 \citep[e.g.][]{Mao2008,Kohri2013,Shima2014}.
Small-scale perturbations generate spectral distortions of CMB by
 Silk dumping due to photon diffusion,
 and future planned experiment PIXIE are also expected to probe the power spectrum up to scales as small as $k=10^4 {\rm Mpc}^{-1}$, and potentially measure $\alpha,\beta$. 
 \citep[]{Chluba:2012we, Powell:2012xz, Chluba:2013pya}. Given the baryonic effects that swamp late time measurements, it seems that such early universe measurements are more promising in that respect.

\section*{Acknowledgments}
We thank Tomoaki Ishiyama and Takahiro Nishimichi for kindly providing us
 their numerical codes.
We also thank Avishai Dekel, Matthias Bartelmann and Chul-Moon Yoo for useful comments.
IBD is supported by the Impuls und Vernetzungsfond of the Helmholtz Association of German Research Centres under grant HZ-NG-603, and the German Science Foundation (DFG) within the Collaborative Research Center 676 ``Particles, Strings and the Early Universe''.
RT was supported in part by JSPS Grant-in-Aid for
Scientific Research (B) (No. 25287062) ``Probing the origin
of primordial mini-halos via gravitational lensing phenomena'', 
 and by Hirosaki University Grant for
 Exploratory Research by Young Scientists.
Numerical computations were carried out on Cray XC30 at Center for
Computational Astrophysics, CfCA, of National Astronomical Observatory
of Japan. 

\bibliographystyle{mn2e} \bibliography{mn-jour,refs-1}

\begin{thebibliography}{75}
\expandafter\ifx\csname natexlab\endcsname\relax\def\natexlab#1{#1}\fi

\bibitem[{Ade {et~al}\mbox{.}(2015)Ade {et~al.}}]{Planck:2015xua}
Ade P., {et~al.}, 2015, arXiv:1502.01589

\bibitem[{{Bagla} \& {Padmanabhan}(1997)}]{bp1997}
{Bagla} J.~S., {Padmanabhan} T., 1997, \mnras, 286, 1023

\bibitem[{{Bagla} \& {Prasad}(2009)}]{bp2009}
{Bagla} J.~S., {Prasad} J., 2009, \mnras, 393, 607

\bibitem[{{Bartelmann} \& {Schneider}(2001)}]{bs01}
{Bartelmann} M., {Schneider} P., 2001, \physrep, 340, 291

\bibitem[{{Ben-Dayan}(2014)}]{Ido2014b}
{Ben-Dayan} I., 2014, arXiv:1408.3004

\bibitem[{Ben-Dayan \& Brustein(2010)}]{BenDayan:2009kv}
Ben-Dayan I., Brustein R., 2010, JCAP, 1009, 007

\bibitem[{{Ben-Dayan} {et~al}\mbox{.}(2013){Ben-Dayan}, {Gasperini}, {Marozzi},
  {Nugier}, \& {Veneziano}}]{ido13}
{Ben-Dayan} I., {Gasperini} M., {Marozzi} G., {Nugier} F., {Veneziano} G.,
  2013, \jcap, 6, 2

\bibitem[{{Ben-Dayan} \& {Kalaydzhyan}(2014)}]{Ido2014}
{Ben-Dayan} I., {Kalaydzhyan} T., 2014, \prd, 90, 083509

\bibitem[{Ben-Dayan {et~al}\mbox{.}(2012)Ben-Dayan, Marozzi, Nugier, \&
  Veneziano}]{BenDayan:2012wi}
Ben-Dayan I., Marozzi G., Nugier F., Veneziano G., 2012, JCAP, 1211, 045

\bibitem[{Betoule {et~al}\mbox{.}(2014)Betoule {et~al.}}]{Betoule:2014frx}
Betoule M., {et~al.}, 2014, Astron.Astrophys., 568, A22

\bibitem[{Bezrukov \& Shaposhnikov(2014)}]{Bezrukov:2014bra}
Bezrukov F., Shaposhnikov M., 2014, Phys.Lett., B734, 249

\bibitem[{{Campbell} {et~al}\mbox{.}(2013){Campbell}, {D'Andrea}, {Nichol},
  {Sako}, {Smith}, {Lampeitl}, {Olmstead}, {Bassett}, {Biswas}, {Brown},
  {Cinabro}, {Dawson}, {Dilday}, {Foley}, {Frieman}, {Garnavich}, {Hlozek},
  {Jha}, {Kuhlmann}, {Kunz}, {Marriner}, {Miquel}, {Richmond}, {Riess},
  {Schneider}, {Sollerman}, {Taylor}, \& {Zhao}}]{Camp2013}
{Campbell} H. {et~al.}, 2013, \apj, 763, 88

\bibitem[{{Casarini} {et~al}\mbox{.}(2011){Casarini}, {Macci{\`o}},
  {Bonometto}, \& {Stinson}}]{casarini2011}
{Casarini} L., {Macci{\`o}} A.~V., {Bonometto} S.~A., {Stinson} G.~S., 2011,
  \mnras, 412, 911

\bibitem[{{Castro} \& {Quartin}(2014)}]{CQ2014}
{Castro} T., {Quartin} M., 2014, \mnras, 443, L6

\bibitem[{Chluba {et~al}\mbox{.}(2012)Chluba, Erickcek, \&
  Ben-Dayan}]{Chluba:2012we}
Chluba J., Erickcek A.~L., Ben-Dayan I., 2012, Astrophys.J., 758, 76

\bibitem[{Chluba \& Jeong(2014)}]{Chluba:2013pya}
Chluba J., Jeong D., 2014, Mon.Not.Roy.Astron.Soc., 438, 2065

\bibitem[{{Crocce} {et~al}\mbox{.}(2006){Crocce}, {Pueblas}, \&
  {Scoccimarro}}]{crocce2006}
{Crocce} M., {Pueblas} S., {Scoccimarro} R., 2006, \mnras, 373, 369

\bibitem[{{Dodelson} \& {Vallinotto}(2006)}]{DV2006}
{Dodelson} S., {Vallinotto} A., 2006, \prd, 74, 063515

\bibitem[{{Fedeli} \& {Moscardini}(2014)}]{Fedeli2014}
{Fedeli} C., {Moscardini} L., 2014, \mnras, 442, 2659

\bibitem[{Hamada {et~al}\mbox{.}(2014)Hamada, Kawai, Oda, \&
  Park}]{Hamada:2014iga}
Hamada Y., Kawai H., Oda K.-y., Park S.~C., 2014, Phys.Rev.Lett., 112, 241301

\bibitem[{{Hamana} \& {Futamase}(2000)}]{HF2000}
{Hamana} T., {Futamase} T., 2000, \apj, 534, 29

\bibitem[{{Harnois-D{\'e}raps} {et~al}\mbox{.}(2014){Harnois-D{\'e}raps}, {van
  Waerbeke}, {Viola}, \& {Heymans}}]{HD2014}
{Harnois-D{\'e}raps} J., {van Waerbeke} L., {Viola} M., {Heymans} C., 2014,
  arXiv:1407.4301

\bibitem[{{Hilbert} {et~al}\mbox{.}(2007){Hilbert}, {White}, {Hartlap}, \&
  {Schneider}}]{hilb07}
{Hilbert} S., {White} S.~D.~M., {Hartlap} J., {Schneider} P., 2007, \mnras,
  382, 121

\bibitem[{{Holz}(1998)}]{Holz1998}
{Holz} D.~E., 1998, \apjl, 506, L1

\bibitem[{{Holz}(2001)}]{Holz2001}
{Holz} D.~E., 2001, \apjl, 556, L71

\bibitem[{{Holz} \& {Linder}(2005)}]{Holz2005}
{Holz} D.~E., {Linder} E.~V., 2005, \apj, 631, 678

\bibitem[{{Inoue} {et~al}\mbox{.}(2015){Inoue}, {Takahashi}, {Takahashi}, \&
  {Ishiyama}}]{itti2014}
{Inoue} K.~T., {Takahashi} R., {Takahashi} T., {Ishiyama} T., 2015, \mnras,
  448, 2704

\bibitem[{{Ishiyama} {et~al}\mbox{.}(2009){Ishiyama}, {Fukushige}, \&
  {Makino}}]{ishiyama2009}
{Ishiyama} T., {Fukushige} T., {Makino} J., 2009, PASJ, 61, 1319

\bibitem[{{Ishiyama} {et~al}\mbox{.}(2012){Ishiyama}, {Nitadori}, \&
  {Makino}}]{ishiyama2012}
{Ishiyama} T., {Nitadori} K., {Makino} J., 2012, Proc. Int. Conf. High
  Performance Computing, Networking, Storage and Analysis, SC '12 (Los
  Alamitos, CA: IEEE Computer Society Press), 5, (arXiv:1211.4406)

\bibitem[{{Iwata} \& {Yoo}(2015)}]{IY2014}
{Iwata} K., {Yoo} C.-M., 2015, EPL (Europhysics Letters), 109, 39001

\bibitem[{{Jing} {et~al}\mbox{.}(2006){Jing}, {Zhang}, {Lin}, {Gao}, \&
  {Springel}}]{jing2006}
{Jing} Y.~P., {Zhang} P., {Lin} W.~P., {Gao} L., {Springel} V., 2006, \apjl,
  640, L119

\bibitem[{{J{\"o}nsson} {et~al}\mbox{.}(2006){J{\"o}nsson}, {Dahl{\'e}n},
  {Goobar}, {Gunnarsson}, {M{\"o}rtsell}, \& {Lee}}]{jonsson2006}
{J{\"o}nsson} J., {Dahl{\'e}n} T., {Goobar} A., {Gunnarsson} C., {M{\"o}rtsell}
  E., {Lee} K., 2006, \apj, 639, 991

\bibitem[{{J{\"o}nsson} {et~al}\mbox{.}(2010){J{\"o}nsson}, {Sullivan}, {Hook},
  {Basa}, {Carlberg}, {Conley}, {Fouchez}, {Howell}, {Perrett}, \&
  {Pritchet}}]{jonsson2010}
{J{\"o}nsson} J. {et~al.}, 2010, \mnras, 405, 535

\bibitem[{{Karpenka} {et~al}\mbox{.}(2013){Karpenka}, {March}, {Feroz}, \&
  {Hobson}}]{Karpenka2013}
{Karpenka} N.~V., {March} M.~C., {Feroz} F., {Hobson} M.~P., 2013, \mnras, 433,
  2693

\bibitem[{{Kohri} {et~al}\mbox{.}(2013){Kohri}, {Oyama}, {Sekiguchi}, \&
  {Takahashi}}]{Kohri2013}
{Kohri} K., {Oyama} Y., {Sekiguchi} T., {Takahashi} T., 2013, \jcap, 10, 65

\bibitem[{{Kronborg} {et~al}\mbox{.}(2010){Kronborg}, {Hardin}, {Guy},
  {Astier}, {Balland}, {Basa}, {Carlberg}, {Conley}, {Fouchez}, {Hook},
  {Howell}, {J{\"o}nsson}, {Pain}, {Pedersen}, {Perrett}, {Pritchet},
  {Regnault}, {Rich}, {Sullivan}, {Palanque-Delabrouille}, \&
  {Ruhlmann-Kleider}}]{Kronborg2010}
{Kronborg} T. {et~al.}, 2010, \aap, 514, A44

\bibitem[{{Lewis} {et~al}\mbox{.}(2000){Lewis}, {Challinor}, \&
  {Lasenby}}]{camb}
{Lewis} A., {Challinor} A., {Lasenby} A., 2000, \apj, 538, 473

\bibitem[{{Little} {et~al}\mbox{.}(1991){Little}, {Weinberg}, \&
  {Park}}]{lwp1991}
{Little} B., {Weinberg} D.~H., {Park} C., 1991, \mnras, 253, 295

\bibitem[{{Mao} {et~al}\mbox{.}(2008){Mao}, {Tegmark}, {McQuinn},
  {Zaldarriaga}, \& {Zahn}}]{Mao2008}
{Mao} Y., {Tegmark} M., {McQuinn} M., {Zaldarriaga} M., {Zahn} O., 2008, \prd,
  78, 023529

\bibitem[{{Marra} {et~al}\mbox{.}(2013){Marra}, {Quartin}, \&
  {Amendola}}]{Marra2013}
{Marra} V., {Quartin} M., {Amendola} L., 2013, \prd, 88, 063004

\bibitem[{{Metcalf}(1999)}]{Metcalf1999b}
{Metcalf} R.~B., 1999, \mnras, 305, 746

\bibitem[{{Metcalf} \& {Silk}(1999)}]{metcalf1999}
{Metcalf} R.~B., {Silk} J., 1999, \apjl, 519, L1

\bibitem[{{Metcalf} \& {Silk}(2007)}]{Metcalf2007}
{Metcalf} R.~B., {Silk} J., 2007, Physical Review Letters, 98, 071302

\bibitem[{Minty {et~al}\mbox{.}(2002)Minty, Heavens, \& Hawkins}]{Minty:2001jg}
Minty E.~M., Heavens A.~F., Hawkins M.~R., 2002, Mon.Not.Roy.Astron.Soc., 330,
  378

\bibitem[{{M{\"o}rtsell} {et~al}\mbox{.}(2001){M{\"o}rtsell}, {Goobar}, \&
  {Bergstr{\"o}m}}]{mortsell2001}
{M{\"o}rtsell} E., {Goobar} A., {Bergstr{\"o}m} L., 2001, \apj, 559, 53

\bibitem[{{Neyrinck} \& {Yang}(2013)}]{ny2013}
{Neyrinck} M.~C., {Yang} L.~F., 2013, \mnras, 433, 1628

\bibitem[{{Nishimichi} {et~al}\mbox{.}(2014){Nishimichi}, {Bernardeau}, \&
  {Taruya}}]{nishimichi2014}
{Nishimichi} T., {Bernardeau} F., {Taruya} A., 2014, arXiv:1411.2970

\bibitem[{{Nishimichi} {et~al}\mbox{.}(2009){Nishimichi}, {Shirata}, {Taruya},
  {Yahata}, {Saito}, {Suto}, {Takahashi}, {Yoshida}, {Matsubara}, {Sugiyama},
  {Kayo}, {Jing}, \& {Yoshikawa}}]{nishimichi2009}
{Nishimichi} T. {et~al.}, 2009, \pasj, 61, 321

\bibitem[{{Padmanabhan} \& {Ray}(2006)}]{pr2006}
{Padmanabhan} T., {Ray} S., 2006, \mnras, 372, L53

\bibitem[{{Perlmutter} {et~al}\mbox{.}(1999){Perlmutter}, {Aldering},
  {Goldhaber}, {Knop}, {Nugent}, {Castro}, {Deustua}, {Fabbro}, {Goobar},
  {Groom}, {Hook}, {Kim}, {Kim}, {Lee}, {Nunes}, {Pain}, {Pennypacker},
  {Quimby}, {Lidman}, {Ellis}, {Irwin}, {McMahon}, {Ruiz-Lapuente}, {Walton},
  {Schaefer}, {Boyle}, {Filippenko}, {Matheson}, {Fruchter}, {Panagia},
  {Newberg}, {Couch}, \& {Project}}]{Perl1999}
{Perlmutter} S. {et~al.}, 1999, \apj, 517, 565

\bibitem[{{Planck Collaboration} {et~al}\mbox{.}(2014{\natexlab{a}}){Planck
  Collaboration}, {Ade}, {Aghanim}, {Armitage-Caplan}, {Arnaud}, {Ashdown},
  {Atrio-Barandela}, {Aumont}, {Baccigalupi}, {Banday}, \&
  et~al.}]{planck2013xvi}
{Planck Collaboration} {et~al.}, 2014{\natexlab{a}}, \aap, 571, A16

\bibitem[{{Planck Collaboration} {et~al}\mbox{.}(2014{\natexlab{b}}){Planck
  Collaboration}, {Ade}, {Aghanim}, {Armitage-Caplan}, {Arnaud}, {Ashdown},
  {Atrio-Barandela}, {Aumont}, {Baccigalupi}, {Banday}, \&
  et~al.}]{planck2013xxii}
{Planck Collaboration} {et~al.}, 2014{\natexlab{b}}, \aap, 571, A22

\bibitem[{Powell(2012)}]{Powell:2012xz}
Powell B.~A., 2012, arXiv:1209.2024

\bibitem[{{Quartin} {et~al}\mbox{.}(2014){Quartin}, {Marra}, \&
  {Amendola}}]{QMA2014}
{Quartin} M., {Marra} V., {Amendola} L., 2014, \prd, 89, 023009

\bibitem[{{Rauch}(1991)}]{Rauch1991}
{Rauch} K.~P., 1991, \apj, 374, 83

\bibitem[{{Riess} {et~al}\mbox{.}(1998){Riess}, {Filippenko}, {Challis},
  {Clocchiatti}, {Diercks}, {Garnavich}, {Gilliland}, {Hogan}, {Jha},
  {Kirshner}, {Leibundgut}, {Phillips}, {Reiss}, {Schmidt}, {Schommer},
  {Smith}, {Spyromilio}, {Stubbs}, {Suntzeff}, \& {Tonry}}]{Riess1998}
{Riess} A.~G. {et~al.}, 1998, \aj, 116, 1009

\bibitem[{{Riess} {et~al}\mbox{.}(2007){Riess}, {Strolger}, {Casertano},
  {Ferguson}, {Mobasher}, {Gold}, {Challis}, {Filippenko}, {Jha}, {Li},
  {Tonry}, {Foley}, {Kirshner}, {Dickinson}, {MacDonald}, {Eisenstein},
  {Livio}, {Younger}, {Xu}, {Dahl{\'e}n}, \& {Stern}}]{Riess2007}
{Riess} A.~G. {et~al.}, 2007, \apj, 659, 98

\bibitem[{{Rudd} {et~al}\mbox{.}(2008){Rudd}, {Zentner}, \&
  {Kravtsov}}]{rudd2008}
{Rudd} D.~H., {Zentner} A.~R., {Kravtsov} A.~V., 2008, \apj, 672, 19

\bibitem[{{Schaye} {et~al}\mbox{.}(2010){Schaye}, {Dalla Vecchia}, {Booth},
  {Wiersma}, {Theuns}, {Haas}, {Bertone}, {Duffy}, {McCarthy}, \& {van de
  Voort}}]{schaye2010}
{Schaye} J. {et~al.}, 2010, \mnras, 402, 1536

\bibitem[{{Shimabukuro} {et~al}\mbox{.}(2014){Shimabukuro}, {Ichiki}, {Inoue},
  \& {Yokoyama}}]{Shima2014}
{Shimabukuro} H., {Ichiki} K., {Inoue} S., {Yokoyama} S., 2014, \prd, 90,
  083003

\bibitem[{{Smith} {et~al}\mbox{.}(2014){Smith}, {Bacon}, {Nichol}, {Campbell},
  {Clarkson}, {Maartens}, {D'Andrea}, {Bassett}, {Cinabro}, {Finley},
  {Frieman}, {Galbany}, {Garnavich}, {Olmstead}, {Schneider}, {Shapiro}, \&
  {Sollerman}}]{smith2014}
{Smith} M. {et~al.}, 2014, \apj, 780, 24

\bibitem[{{Smith} \& {Markovic}(2011)}]{sm2011}
{Smith} R.~E., {Markovic} K., 2011, Physical Review D, 84, 063507

\bibitem[{{Smith} {et~al}\mbox{.}(2003){Smith}, {Peacock}, {Jenkins}, {White},
  {Frenk}, {Pearce}, {Thomas}, {Efstathiou}, \& {Couchman}}]{smith2003}
{Smith} R.~E. {et~al.}, 2003, \mnras, 341, 1311

\bibitem[{{Springel}(2005)}]{springel2005}
{Springel} V., 2005, \mnras, 364, 1105

\bibitem[{{Sullivan} {et~al}\mbox{.}(2011){Sullivan}, {Guy}, {Conley},
  {Regnault}, {Astier}, {Balland}, {Basa}, {Carlberg}, {Fouchez}, {Hardin},
  {Hook}, {Howell}, {Pain}, {Palanque-Delabrouille}, {Perrett}, {Pritchet},
  {Rich}, {Ruhlmann-Kleider}, {Balam}, {Baumont}, {Ellis}, {Fabbro},
  {Fakhouri}, {Fourmanoit}, {Gonz{\'a}lez-Gait{\'a}n}, {Graham}, {Hudson},
  {Hsiao}, {Kronborg}, {Lidman}, {Mourao}, {Neill}, {Perlmutter}, {Ripoche},
  {Suzuki}, \& {Walker}}]{Sull2011}
{Sullivan} M. {et~al.}, 2011, \apj, 737, 102

\bibitem[{{Takahashi} {et~al}\mbox{.}(2011){Takahashi}, {Oguri}, {Sato}, \&
  {Hamana}}]{taka11b}
{Takahashi} R., {Oguri} M., {Sato} M., {Hamana} T., 2011, \apj, 742, 15

\bibitem[{{Takahashi} {et~al}\mbox{.}(2012){Takahashi}, {Sato}, {Nishimichi},
  {Taruya}, \& {Oguri}}]{takahashi2012}
{Takahashi} R., {Sato} M., {Nishimichi} T., {Taruya} A., {Oguri} M., 2012,
  \apj, 761, 152

\bibitem[{{Valageas}(2000)}]{Valageas2000}
{Valageas} P., 2000, \aap, 354, 767

\bibitem[{{van Daalen} {et~al}\mbox{.}(2011){van Daalen}, {Schaye}, {Booth}, \&
  {Dalla Vecchia}}]{vanDaalen2011}
{van Daalen} M.~P., {Schaye} J., {Booth} C.~M., {Dalla Vecchia} C., 2011,
  \mnras, 415, 3649

\bibitem[{{Viel} {et~al}\mbox{.}(2012){Viel}, {Markovic}, {Baldi}, \&
  {Weller}}]{viel2012}
{Viel} M., {Markovic} K., {Baldi} M., {Weller} J., 2012, \mnras, 421, 50

\bibitem[{{Vogelsberger} {et~al}\mbox{.}(2014{\natexlab{a}}){Vogelsberger},
  {Genel}, {Springel}, {Torrey}, {Sijacki}, {Xu}, {Snyder}, {Bird}, {Nelson},
  \& {Hernquist}}]{vogels2014n}
{Vogelsberger} M. {et~al.}, 2014{\natexlab{a}}, \nat, 509, 177

\bibitem[{{Vogelsberger} {et~al}\mbox{.}(2014{\natexlab{b}}){Vogelsberger},
  {Genel}, {Springel}, {Torrey}, {Sijacki}, {Xu}, {Snyder}, {Nelson}, \&
  {Hernquist}}]{vogels2014}
{Vogelsberger} M. {et~al.}, 2014{\natexlab{b}}, \mnras, 444, 1518

\bibitem[{{Weinberg} {et~al}\mbox{.}(2013){Weinberg}, {Mortonson},
  {Eisenstein}, {Hirata}, {Riess}, \& {Rozo}}]{weinberg2013}
{Weinberg} D.~H., {Mortonson} M.~J., {Eisenstein} D.~J., {Hirata} C., {Riess}
  A.~G., {Rozo} E., 2013, \physrep, 530, 87

\bibitem[{{White} \& {Croft}(2000)}]{wc2000}
{White} M., {Croft} R.~A.~C., 2000, \apj, 539, 497

\bibitem[{{Yoo} {et~al}\mbox{.}(2008){Yoo}, {Ishihara}, {Nakao}, \&
  {Tagoshi}}]{Yoo2008}
{Yoo} C., {Ishihara} H., {Nakao} K., {Tagoshi} H., 2008, Progress of
  Theoretical Physics, 120, 961

\end{thebibliography}

\appendix

\section{Fitting formula for non-linear matter power spectra}

This Appendix presents our fitting formula for the matter power
 spectra with running and running of running of the spectral index.
Our formula is based on the halofit model
 \citep[][hereafter T12]{smith2003,takahashi2012},
 but slightly modified for the running models.

To find the best fitting parameters in the halofit model, we use the
 standard chi-square analysis, which is defined as
\begin{equation}
 \chi^2 = \sum_{i} \sum_{z=0}^{1.5} \sum_{k=k_{\rm min}}^{k_{\rm max}}
 \frac{\left[ P_{i, {\rm model}}(k,z)-P_{i, {\rm sim}}(k,z) \right]^2}
 {P_{i, {\rm sim}}(k,z)^2},
\end{equation}
where $P_{i, {\rm model}}$ is the power spectrum in the theoretical model,
 $P_{i, {\rm sim}}$ is those in the simulation results, and $i$ denotes the
 seven models of the running parameters $(\alpha,\beta)$.
The $\chi^2$ is summed over redshifts $z=0, 0.3, 0.6, 1$ and $1.5$.
We use the wavenumber $k$ larger than $2\,h {\rm Mpc}^{-1} (=k_{\rm min})$
 where the Gaussian error of $P(k)$ is less than $1\%$.
We set the maximum wavenumber $k_{\rm max}$ so that the measured power
 spectrum is much larger ($10$ times larger) than the shot noise.
Thus, $k_{\rm max}$ depends on the parameters of $\alpha, \beta$ and $z$.
The maximum (minimum) $k_{\rm max}$ is $600 (100) h{\rm Mpc}^{-1}$ for
 $\alpha=\beta=0.06 (0)$ at $z=0 (1.5)$.

First, we fit the simulation results for the no running model
 ($\alpha=\beta=0$).
Here, we slightly modify the formula in T12.
The halofit in T12 can be used up to $k<30h {\rm Mpc}^{-1}$
 within $10\%$ error, but it asymptotically overestimate the $P_{\rm m}(k)$
 at very large $k (>100h {\rm Mpc}^{-1})$.
The fitting parameters are the same as in T12,
 except for the following two parameters:
\begin{eqnarray}
 && \hspace{-0.5cm} \log_{10} c_{\rm n} = -0.3370 + 2.1842 n_{\rm eff}
 + 0.9604 n_{\rm eff}^2 + 2.3764 C,
 \nonumber  \\
 && \hspace{-0.5cm} \gamma_{\rm n} = 0.1027 - 0.0219 n_{\rm eff}
 + 2.6785 C 
 \nonumber  \\
 && \hspace{0.5cm} - \left( 0.2866 + 0.1206 n_{\rm eff} \right)
 \ln \left( \frac{k}{h {\rm Mpc}^{-1}} \right).
\label{halofit_params}
\end{eqnarray}
In the last term, if one uses the wavenumber in unit of ${\rm Mpc}^{-1}$ 
 (not $h {\rm Mpc}^{-1}$), then one has to insert a specific value of $h$.
The equation (\ref{halofit_params}) gives a minor correction to the 
 formula in T12, which corrects the overestimation in T12 at
 $k > 100h {\rm Mpc}^{-1}$. 
The $P_{\rm m}(k)$ with and without the correction in equation
 (\ref{halofit_params}) are different within $17\%$ for $k<100h
 {\rm Mpc}^{-1}$.

The ratios of the power spectra with the running to that without it
 are fitted as,
\BE
 \frac{P(k,z;\alpha,\beta)}{P(k,z;\alpha=\beta=0)} =
 \left[ 1 + 0.123 \left( \frac{k}{h {\rm Mpc}^{-1}} \right)^{0.794}
 \right]^{p(z;\alpha,\beta)},
\label{pk_ratio}
\EE
with
\BE
 p(z;\alpha,\beta)=\left( 3.32\alpha + 4.72 \beta \right)
 \left( 1+z \right)^{0.379}
\label{eq_f_p}
\EE
The RHS of equation (\ref{pk_ratio}) corresponds to an enhancement factor.
Note that the parameters $n_{\rm eff}$ and $C$ in equation
 (\ref{halofit_params}) are evaluated in the no running model even when
 computing $P_{\rm m}(k)$ for the running models.
Using our fitting formula, the simulation results can be reproduced within a
 relative error of $18.3 (28.3) \%$ for $k<300 (600) h {\rm Mpc}^{-1}$ at $z=0-1.5$. 
The root-mean-square deviation between the
 theoretical model and the simulation results is about $5.1 \%$.

\begin{figure*}
\vspace*{1.0cm}
\includegraphics[width=150mm]{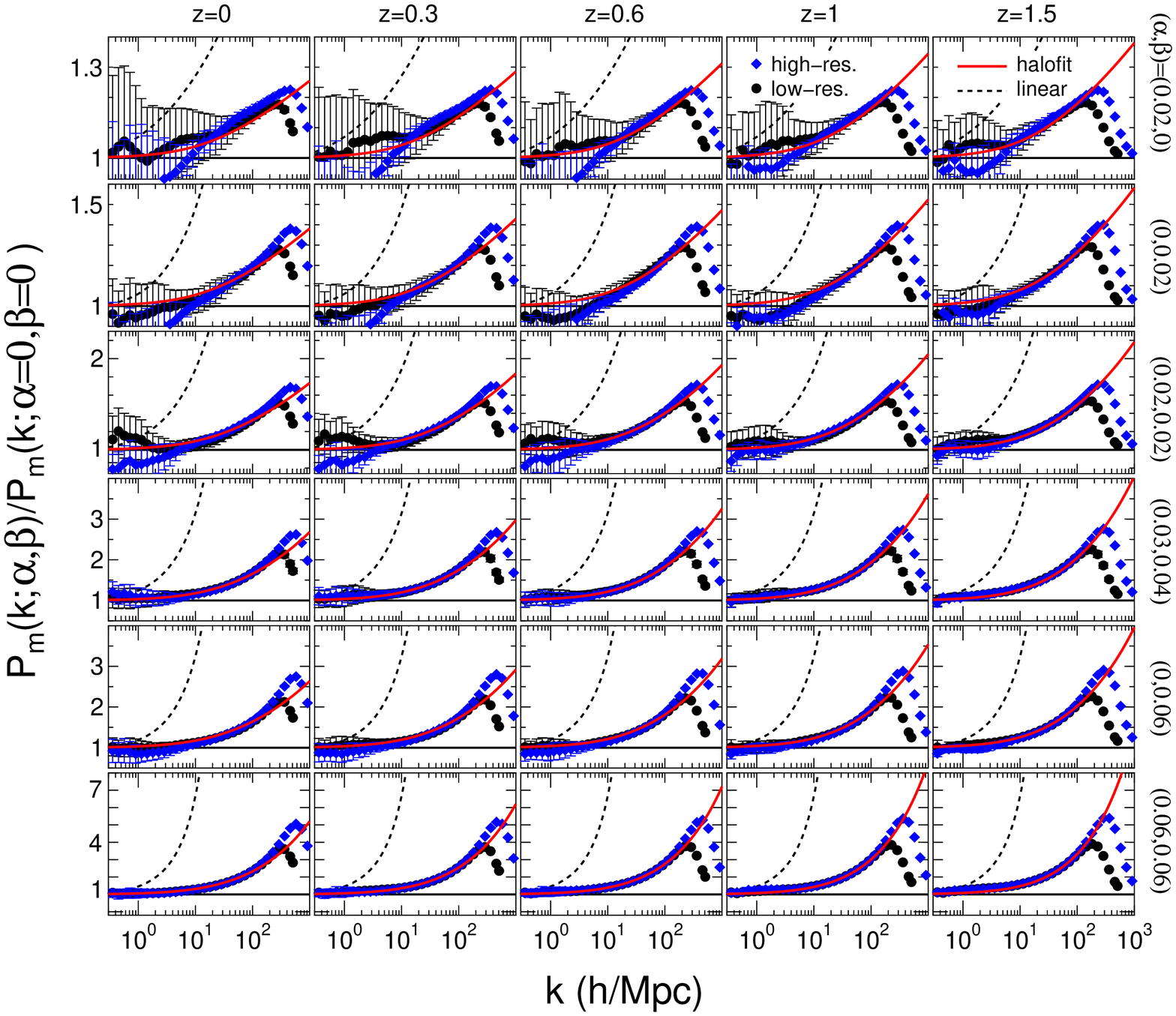}
\caption{
Ratio of the power spectra with the
 running $P_{\rm m}(k,z;\alpha,\beta)$ to that without the running
 $P_{\rm m}(k,z;\alpha=0,\beta=0)$.
The blue diamonds (black dots) are the ratios in the high-(low-)resolution
 simulations.
The dotted curves are the ratios in the linear power spectra. 
The red solid curves are our enhancement factor in equation (\ref{pk_ratio}).
The small dumping at very high $k (>100h {\rm Mpc}^{-1})$ of the simulation
 results are due to the shot noise in the denominator
 $P_{\rm m}(k,z;\alpha=0,\beta=0)$.
}
\label{fig_pk_ratio}
\vspace*{0.5cm}
\end{figure*}

Fig.\ref{fig_pk_ratio} shows the ratio of the power spectrum with the
 running, $P_{\rm m}(k,z;\alpha,\beta)$, to that without it,
 $P_{\rm m}(k,z;\alpha=0,\beta=0)$.
The red solid curves show the enhancement of power spectra described by  
 our fitting factor in Eq.(\ref{pk_ratio}), which
 reproduces our simulation results very well.

\end{document}